\documentclass[aps,pre,twocolumn,superscriptaddress,reprint]{revtex4-1}
\usepackage{graphicx, color}
\usepackage{amsmath}
\usepackage{times}
\def\beq{\begin{equation}}
\def\eeq{\end{equation}}
\def\bea{\begin{eqnarray}}
\def\eea{\end{eqnarray}}
\newcommand{\eref}[1]{Eq.~\eqref{#1}}

\begin{document}

\title{Generating true minima in constrained variational formulations via modified Lagrange multipliers}
\author{Francisco J. Solis}
\email{francisco.solis@asu.edu}
\affiliation{School of Mathematical and Natural Sciences, Arizona State University, Glendale, Arizona, 85306, USA}
\author{Vikram Jadhao}
\affiliation{Department of Materials Science and Engineering, Northwestern University, Evanston, Illinois 60208, USA}
\author{Monica Olvera de la Cruz}
\email{m-olvera@northwestern.edu}
\affiliation{Department of Materials Science and Engineering, Northwestern University, Evanston, Illinois 60208, USA}

\begin{abstract}
Variational principles are important in the investigation of large classes of physical systems. They can be used both as analytical methods as well as starting points for the formulation of powerful computational techniques such as dynamical optimization methods. Systems with charged objects in dielectric media and systems with magnetically active particles are important examples. In these examples and other important cases, the variational principles describing the system are required to obey a number of constraints. These constraints are implemented within the variational formulation by means of Lagrange multipliers. Such constrained variational formulations are in general not unique. For
the application of efficient simulation methods, one must find specific formulations that satisfy a number of important conditions. An often required condition is that the functional be positive-definite, in other words, its extrema be actual minima. In this article, we present a general approach to attack the problem of finding, among equivalent variational functionals, those that generate true minima. The method is based on the modification of the Lagrange multiplier which allows us to generate large families of effective variational formulations associated with a single original constrained variational principle. We demonstrate its application to different examples and, in particular, to the important cases
of Poisson and Poisson-Boltzmann equations.  We show how to obtain variational formulations for these systems with extrema that are always minima.
\end{abstract}

\maketitle

\section{Introduction}
The formulation of equilibrium or dynamical problems in terms of variational
principles is important in several different contexts. Standard formulations
of the laws of mechanics, thermodynamics, electromagnetism and other
classical theories, all use variational principles \cite{landau,schwinger,landau-ctf}. Beyond their aesthetic appeal, these variational principles are used in practice as ways to investigate the stability of physical phenomena, where deviations from equilibrium states or trajectories is important. Path integral
approaches to quantization of these systems \cite{henneaux,tyutin} also require expression of the properties of dynamical systems away from their classical equilibrium trajectories as it is understood that quantum particles explore the
full space of possible trajectories during their evolution. Finally, the applications of variational principles have emerged as key components of numerical methods for the investigation of complex systems. These variational functionals allow, for example, the determination of equilibrium conditions using direct minimization methods. Molecular dynamics simulations benefit of variational formulations as well: it is sometimes possible to replace the exact evaluation of internal variables for an approximate solution dictated by the given variational principle \cite{car-parrinello,pasichnyk,rottler-maggs,jso1}.

This last type of application and, in particular, the implementation of approximate molecular dynamics schemes for simulation of charged systems, is the key motivation of this work. During the simulation of such systems, it appears necessary to evaluate the electric field (or the electric potential) produced by the mobile charges at each time step, in order to determine the forces acting on the charges. This intermediate step consumes considerable computational resources and alternatives to this procedure are highly desirable. Techniques motivated by the idea of a dynamical optimization of the functional \cite{car-parrinello}, for example, replace the computation of the exact potential at each step for an approximate value. Consider the problem of simulating charges in the presence of dielectric heterogeneities. The calculation of total electrostatic potential in such a system requires the computation of the induced polarization charge. Instead of an exact calculation, the polarization 
charge is promoted to a (fictitious) dynamical variable with small mass. The potential contribution of the dynamically evolving polarization charge density is easier to compute, while the (fictitious) dynamics of the induced charge density is also quickly determined \cite{jso1}. For such a scheme to work, however, it is necessary to produce a functional with a number of key properties.

For a given physical system, there is an infinite number of possible variational principles that reproduce the equilibrium conditions or dynamical equations of the system. But not all variational formulations are equally useful. We spell below in detail the necessary properties of a useful functional. It is clear, however, that a method to generate different variational functionals is helpful in finding functionals suitable for applications. We focus on specific type of functionals where not only an extremum is sought, but a number of constraints on
the functional variables are imposed through Lagrange multipliers. As we show in our examples, the family of systems described by such functionals is large and contains examples of both fundamental and practical relevance.

This article presents a method for generating large families of variational functionals for a single specific problem. Our formalism is developed for problems that satisfy constraints. We elaborate on the structure of the first and second variations of the functional with respect to the base function variable and the Lagrange multiplier. We finally show its application to several concrete examples, including several important problems in the context of charged systems. While previously we have carried out this program in the context of two important examples \cite{jso1,jso2,jso3}, generalizations of the implicit method were not obvious. This article presents the methods employed there in a broad and more formal context.

In section II, we establish important characteristics of useful functionals. In section III, we examine the variations of the simplest general case of a functional with an unmodified constraint. In section IV, we introduce the general method for the construction of new functionals and derive the properties of their variations. Finally, in section V, we present in detail a large number of examples demonstrating the usefulness of the new method. We close with some future perspectives in section VI. 

\section{Properties of a useful functional}
We begin by setting some notations. We denote a variational functional as $I[\rho]$.  To avoid unwieldy notation, we stipulate that the functional variable $\rho$ comprises the field $\rho({x})$ itself, along with a number of its partial derivatives $\partial_{i}\rho$, and that the functional is presented as an integral over space:
\beq
I[\rho]=\int_{V}f(\rho,\partial_{i}\rho,\ldots).
\eeq
Here, $f$ is the functional density. To simplify notation, we will omit the variable of integration in most expressions but will restore it where it might help clarify the content of an expression. The reduced notation $I[\rho]$ instead of a more explicit $I[\rho,\ \partial_{i}\rho,\ldots]$, should also cause  no confusion. We also note that the base variables can be a vector of variables, but we will not write them explicitly unless it is necessary. To further maintain a streamlined notation, we will assume that suitable boundary conditions are set so as to permit all necessary by-parts integrations such that the boundary terms in 
our expressions vanish.

Let us introduce the general form of a functional of a system with a constraint:
\begin{equation}\label{eq:CF}
I[\rho;\psi]=I_{o}[\rho]-I_{c}[\rho;\psi],
\end{equation}
with the constraint part taking the explicit form: 
\beq
I_{c}[\rho;\psi]=\int\psi C[\rho].
\eeq
That is, $\psi$ is a Lagrange multiplier enforcing the constraint $C[\rho]=0$. In writing $I[\rho;\psi]$, we wish to distinguish the role of the multiplier $\psi.$ $I_{o}$ is the part of the functional independent of the constraint and one that does not depend on the multiplier. The conditions of equilibrium for the system are a set of equations for $\rho$ and $\psi$. We denote a solution of the system as $(\rho^{*};\psi^{*}).$

It is clear from Eq.~\eqref{eq:CF} that the constrained functional $I$ is a functional of two variable fields. From the viewpoint of applications, it is desirable to construct a functional with a single variational field. In other words, we seek the elimination of one of the two variables $\rho$ or $\psi$, and the following discussion regarding the properties that we associate with a useful functional should be viewed in the light of this goal.

We identify four important properties of variational functionals. These are: (\textbf{A}) Physical extremal points, (\textbf{B}) True equilibrium evaluation, 
(\textbf{C}) Convexity, and (\textbf{D}) Simplicity. A functional that satisfies properties (\textbf{A}), (\textbf{B}), and (\textbf{C}) leads to a variational principle that generates, as its extrema, the true minima associated with the problem.

Property (\textbf{A}) of our functional simply requires that at least some of the stationary points of the functional should correspond to physically relevant
configurations of the field. The stationarity condition will be written as:
\beq
\frac{\delta I[\rho]}{\delta\rho}=Q(\rho)=0.
\eeq
This expression defines the \emph{equilibrium equation}.
It typically involves derivatives of the field and should be recognizable as the expected condition of equilibrium for the system. Note that, often, we will  write the functional derivative as $\delta_{\rho} I$. Also, note that the result of the functional differentiation is a function of the space variable $x$ on which the field variable $\rho(x)$ depends. Where it does not cause confusion, we omit explicitly mentioning the space variable. 

Even when the functional might reproduce, upon variation, the equation of motion, its value at that point can be in itself meaningless. In most cases of interest, the functional can be identified with the energy of the system, the action of a trajectory, or some other suitable thermodynamic potential. We note that this simple natural condition is not satisfied in a number of well studied cases: some formulations of electrostatics employ functionals that have the property of evaluating to the negative of the true electrostatic energy at their equilibrium points \cite{allen1,jackson}.

Let us consider the case of a constrained functional. As the conditions of equilibrium include the satisfaction of the constraint $C(\rho)=0$ , we note that evaluation of the functional at equilibrium reduces to the evaluation of just the first term on the right hand side of Eq.~\eqref{eq:CF}:
\beq
I[\rho^{*};\psi^{*}]=I_{o}[\rho^{*}]-I_{c}[\rho^{*};\psi^{*}]=I_{o}[\rho^{*}].
\eeq
In general, we make our property (\textbf{B}) to require that evaluation of the functional be meaningful at equilibrium, producing physically interpretable values of energy, action or thermodynamic potentials. In the case of a constrained functional, we see that this is a property of the unconstrained term $I_{o}[\rho]$. 

Our third property (\textbf{C}) is convexity or positive-definiteness. Specifically, we mean that the second variation of the functional at its extremum point be strictly positive, thus implying that the point of extremum is a minimum of the functional. In practice, this property is crucial for applications but might also be difficult to obtain. All gradient methods to find stationary points as well as the development of dynamical simulation schemes
require that the functional considered be a minimum at its stationary point. This condition can be schematically stated in terms of the second variation of the functional with respect to the field, 
\beq\label{eq:positive_svar}
\delta^{2}I[\rho]|_{\rho = \rho^{*}}>0.
\eeq
In concrete examples, the second functional derivative gives rise to a distribution (a generalized function containing delta functions and its derivatives). The positivity implied in Eq.~\eqref{eq:positive_svar} is to be understood as the integration of such an expression against two equal functions in the space of allowed variations. Thus, the above condition is in fact a shorthand for the more explicit statement:
\begin{equation}\label{eq:secondvar}
\delta^{2}I[\rho] = \int dx\int dy\left[\delta\rho(x)\frac{\delta^{2}I}{\delta\rho(x)\delta\rho(y)}\delta\rho(y)\right]>0,
\end{equation}
where $\delta\rho$ is a non-zero arbitrary variation. As with property (\textbf{B}), we actually only demand this condition at the stationary point $\rho^{*}$. We expect, in general, that there be a finite-size neighborhood of the stationary point in which the condition is still valid. We will often refer to the functionals that preserve the convexity property as positive-definite or convex functionals. Similarly, functionals that result in a strictly negative second variation at the extremum will be termed as negative-definite or concave functionals. We note that the standard functional of electrostatic potential for Poisson-Boltzmann theory is a concave functional \cite{radke,fogolari} and so is the functional found in textbooks for the case of electrostatics \cite{schwinger}. In a later section, we will show how to transform these functionals into convex functionals using the method presented here. 

Our final property (\textbf{D}) is that of simplicity, which is of course fairly subjective. We are interested in obtaining new functionals for the same problem that somehow simplify its handling and/or have sound physical interpretations. Concretely, we seek to have variables in our functional that are as simple as possible. In practice, this can mean, for example, to build our functionals with scalar variables instead of the more complicated geometric objects such as vectors. 

Below, we develop a methodology to modify existing functionals with the aim of satisfying most or all of these properties. We first examine the properties of unmodified functionals and then show a technique for modifications. 

\section{Variations of constrained functionals}\label{sec:stdvar}
In this section, we analyze the properties of a generic constrained functional that is defined using a Lagrange multiplier. We start from the form introduced above in Eq.~\eqref{eq:CF}:
\begin{equation}\label{eq:CF1}
I[\rho;\psi]=I_{o}[\rho]-I_{c}[\rho;\psi]=I_{o}[\rho]-\int\psi C[\rho].
\end{equation}
Variation with respect to the base variable $\rho$ leads to an equilibrium
equation containing the multiplier in linear form: 
\begin{equation}\label{eq:varL}
\delta_{\rho}I[\rho;\psi]=Q(\rho,\psi),
\end{equation}
so that the first condition for determining extrema becomes
\begin{equation}\label{eq:Q}
Q(\rho,\psi)=\frac{\delta I_{o}}{\delta\rho}-\psi\frac{\delta C}{\delta\rho}=0.
\end{equation}
The solution of this equation is, in general, non-trivial as the variation
$\delta C/\delta\rho$ is a distribution containing derivatives of $\delta$ functions, rendering this expression in effect a differential equation for $\psi$. The second condition, obtained from variation of the functional with respect to the multiplier is simply the constraint:
\beq
\delta_{\psi}I[\rho;\psi]=-C[\rho]=0.
\eeq
In fuller notation, this reads:
\begin{eqnarray}
\frac{\delta I[\rho;\psi]}{\delta\psi(x)}
&=&-\int dy\frac{\delta\psi(y)}{\delta\psi(x)}C[\rho(y)]\\ \nonumber
&=&-\int dy\delta(y-x)C[\rho(y)]=-C[\rho(x)]=0.\nonumber
\end{eqnarray}
We will omit most such explicit derivations. 

Together, the pair of equations $Q(\rho,\psi)=0$, and $C[\rho]=0$, determine the solutions to our system, $(\rho^{*},\psi^{*})$. Motivated by the goal of expressing the functional in terms of a single variational field, we adopt a slightly different route. Before taking variations with respect to the multiplier, we first solve the equation $Q(\rho,\psi)=0.$

The nature of the solutions of the relation $Q(\rho,\psi)=0$ is of course highly dependent on the system considered. In some generality, we expect that for a given value of $\psi$ there be a value of $\rho$ that satisfies the given equation. We therefore define a functional $R[\psi]$ as this precise value. That is:
\begin{equation}\label{eq:Q=0}
Q(R[\psi],\psi)=0.
\end{equation}

In some cases, we might wish to retain $\rho$ as the variable and eliminate $\psi$. As, in general, there might not be solutions to the equation with a prescribed value of $\rho$, we must restrict this variable to the range of the functional $R$. The inverse of $R$, if exists, will be denoted $S[\rho]$. Constructing functionals with either function variables ($\rho$ or $\psi$) is of practical importance, and we will write down results relevant to both formats. To start, we consider the case where we continue to use $\psi$ as the sole variational field. The constraint becomes
\beq\label{eq:D}
D[\psi]=C[R[\psi]].
\eeq
Substituting the solution $\rho = R[\psi]$ in Eq.~\eqref{eq:CF1}, we obtain a restricted evaluation $I_{\psi}$ of the functional $I$: 
\beq\label{eq:Ipsi}
I_{\psi}[\psi]=I_{o}[R[\psi]]-I_{c}[R[\psi];\psi]=I_{o}[R[\psi]]-\int\psi D[\psi].
\eeq

We can now search for the extrema of this functional by again taking
variations of $I_{\psi}$ with respect to $\psi$. In doing so we take advantage
of the form obtained after the variation of $I[\rho;\psi]$, namely, Eq.~\eqref{eq:Q}. 
We have
\begin{eqnarray}\label{eq:varLpsi}
\delta_{\psi}I_{\psi}[\psi]&=&\int Q(R[\psi],\psi)\delta_{\psi}R[\psi]-\int(\delta_{\psi}\psi)D[\psi]\\\nonumber
&=&-D[\psi].
\end{eqnarray}
In this expression $Q$ evaluates to zero by the definition of $R[\psi]$. As a result, we see that a general solution $\psi^{*}$ to this extremization problem is obtained when  
\begin{equation}\label{eq:eqmIpsi}
D[\psi^{*}]=C[R[\psi^{*}]]=0.
\end{equation}
In other words, the extremal condition for the reduced functional $I_{\psi}$ is simply the constraint condition. The solution to the original problem posed in terms of the Lagrange multiplier is obtained from the pair $(\rho^{*}=R[\psi^{*}];\psi^{*})$. 
In typical problems, the solution to the constraint condition will correspond to a physical solution of the problem, and in this sense, the reduced functional
$I_{\psi}$ preserves property (\textbf{A}). Also, it can be checked from Eq.~\eqref{eq:Ipsi} that, owing to Eq.~\eqref{eq:eqmIpsi}, the evaluation of the functional at the extremal point is simply $I_{o}[\psi^{*}]$. Assuming that the latter represents a meaningful quantity, we note that $I_{\psi}$ also preserves property (\textbf{B}).

We can further probe the properties of the reduced functional. To examine its second variation, we first note that $Q(R[\psi],\psi)$ is, by definition, identically zero. Therefore, all of its variations are also identically zero:
\begin{equation}\label{eq:delQ=0}
\delta_{\psi}Q(R[\psi],\psi)=0.
\end{equation}
We use this fact freely below.
From Eq.~\eqref{eq:varLpsi}, we have the second derivative of the functional:
\begin{equation}\label{eq:sfdI}
\delta_{\psi}^{2}I_{\psi}[\psi]= - \int\frac{\delta D[\psi(x')]}{\delta\psi(x)}
\frac{\delta\psi(x')}{\delta\psi(y)} dx',
\end{equation}
where $x$, $y$ are the variables used in the variations $\delta\psi(x)$, $\delta\psi(y)$.
Using the above result in Eq.~\eqref{eq:secondvar} gives the expression for the 
second variation of $I$:
\begin{equation}\label{eq:svarI}
\delta^{2}I_{\psi}[\psi]= -\int \delta D[\psi(x')]\delta\psi(x')dx' = - \int \delta D[\psi] \, \delta \psi,
\end{equation}
where we use the shorthand
\beq\label{eq:delD}
\delta D[\psi] \equiv \delta D [\psi(x')] = \int \frac{\delta D[\psi(x')]}{\delta\psi(x)} \delta\psi(x) dx,
\eeq
and the second equality in Eq.~\eqref{eq:svarI} is the result of using 
the simplified notation. Further, evaluating at the equilibrium point, we have 
\beq\label{eq:svarIpsi}
\delta^{2}I_{\psi}[\psi^{*}] = - \int \delta D[\psi^{*}] \, \delta \psi.
\eeq
Note that, the expression $\delta D[\psi^{*}]$ stands for $\delta D[\psi]|_{\psi=\psi^{*}}$ and such abbreviated notations will often be used in what follows. Equation \eqref{eq:svarIpsi} provides a very useful and compact expression for the second variation of the functional $I_{\psi}$ at equilibrium. 
In general, this second variation can assume any value, positive or negative and hence the reduced functional $I_{\psi}$ may or may not preserve property (\textbf{C}).

Briefly, we note that if we are able to use $\rho$, or a restricted version of it, as a base variable, we would use the functional 
\beq\label{eq:Irho}
I_{\rho}[\rho] = I[\rho,S[\rho]] = I_{o}[\rho] - \int S[\rho]C[\rho].
\eeq
The above functional can be obtained by making the substitution of $\psi=S[\rho]$ 
in Eq.~\eqref{eq:Ipsi} and using the identity $R[S[\rho]] = \rho$, which is true by construction. In essentially the same way as above, the first variation leads to the equilibrium condition 
\begin{equation}\label{eq:eqmIrho}
\int C[\rho]\delta_{\rho}S[\rho]=0
\end{equation}
which always has a solution $\rho^{*}$ when $C[\rho^{*}]=0$.
The second variation has, at the solution point, the form 
\begin{eqnarray}\label{eq:svarrho}
\delta^{2}I_{\rho}[\rho^{*}]&=&-\int\delta S[\rho^{*}(x')]\delta C[\rho^{*}(x')] dx'\\\nonumber
&=& -\int\delta S[\rho^{*}]\delta C[\rho^{*}]
\end{eqnarray}
where we employ the shorthand notations	
\beq\label{eq:delS}
\delta S[\rho] \equiv \delta S[\rho(x')] = \int \frac{\delta S[\rho(x')]}{\delta\rho(x)} \delta\rho(x) dx,
\eeq
and 
\beq\label{eq:delC}
\delta C[\rho] \equiv \delta C[\rho(x')] = \int \frac{\delta C[\rho(x')]}{\delta\rho(y)} \delta\rho(y) dy.
\eeq

Before concluding this section we make a few remarks about the change
of base function variables. We have explicitly written expressions that use
either one of the original base variables, thus writing $I_{\rho}$ or $I_{\psi}.$
To some extent this is simply a change of variables but we must be careful
when the relation between the variables is not one-to-one. If this condition holds, and a one-to-one relation exists, the path followed to obtain
the results using $\rho$ as a variable, can be copied using any other
variable. That is, any such change of variable essentially leads
to similar results. In particular, we note that the second variation
formula we have derived is the product of two first order variations.
Therefore, a change in variables modifies the result by the multiplication
of two similar factors arising from the chain rule. The change of
variables, unless singular, does not modify the sign of the second variation. 

\section{The method of modified Lagrange multipliers}\label{sec:modvar}
We now consider a general method to obtain a new functional that has the same set of extremal points as the original reduced functional. The key idea is to replace the multiplier $\psi$ by a different function that reduces to the same value when the constraint $C$ is satisfied. We show that the extrema of the new functional coincide with those of the original but that this change might improve the
properties of the functional by, for example, making it positive-definite. That is, we conserve properties (\textbf{A}) and (\textbf{B}) and improve to a form that also satisfies (\textbf{C}). We develop the formalism using the expressions for the functional in terms of the multiplier $\psi$; that is, we look at modifying the properties of the  functional $I_{\psi}[\psi]$ given by Eq.~\eqref{eq:Ipsi}.

We first remind the reader that the function $R[\psi]$ is defined via Eq.~\eqref{eq:Q=0} and the function $D[\psi]$ is simply the evaluation of the constraint $C$ at the point $\rho = R[\psi]$. We can now proceed to obtain a modified multiplier $\psi_{m}$. A fairly general choice is to write:
\begin{equation}\label{eq:modlm}
\psi_{m}=\Psi[\psi]=h^{-1}[D[\psi]+h(\psi)]
\end{equation}
where $h$ is a function or functional with a well defined inverse and $\Psi$ is a shorthand for the expression on the right. With this choice, we see that 
\begin{equation}\label{eq:equalateqm}
\Psi[\psi^{*}]=\psi^{*}
\end{equation}
since $D[\psi^{*}]=C[R[\psi^{*}]]=0$ and $h^{-1}[h(\psi^{*})] = \psi^{*}$. 
In other words, a function constructed this way replicates the value of the multiplier when the constraint is satisfied. When the condition is not satisfied, the values are in general different. 

We now claim that, it is possible to simply replace the multiplier $\psi$ with this modified multiplier $\psi_{m}$ to obtain the \emph{modified reduced functional} $F_{\psi}$:
\begin{equation}\label{eq:Fpsi}
F_{\psi}[\psi]=I_{o}[R[\psi]]-\int\Psi[\psi]D[\psi].
\end{equation}
It should be clear that when the constraint is satisfied, this functional
still evaluates to the same value as the original reduced functional, that is:
\beq
F_{\psi}[\psi^{*}]=I_{\psi}[\psi^{*}] = I_{o}[R[\psi^{*}]].
\eeq
Thus, we preserve property (\textbf{B}). It is also not hard to see that the
equilibrium equation resulting from this functional is satisfied by the solution $\psi^{*}$ of $D[\psi]=0$. To show this, we first write the functional $F_{\psi}$
in Eq.~\eqref{eq:Fpsi} as 
\begin{align}
F_{\psi}[\psi] &= I_{o}[R[\psi]]-\int\psi D[\psi] - 
\int(\Psi[\psi] - \psi)D[\psi] \\ \notag
&= I_{\psi}[\psi] - \int(\Psi[\psi] - \psi)D[\psi],
\end{align}
where we employed \eref{eq:Ipsi}, that provides the unmodified functional, to obtain the second equality above. 
Variation of the above functional is
\begin{equation}\label{eq:fvar}
\delta_{\psi}F_{\psi} = \delta_{\psi} I_{\psi} - \int \delta_{\psi}(\Psi[\psi] - \psi)D[\psi] - \int(\Psi[\psi] - \psi) \delta_{\psi} D[\psi]
\end{equation}
This expression evaluates to zero when we use the original solution
$\psi^{*}$:
\begin{eqnarray}
\delta_{\psi}F_{\psi}\big\vert_{\psi=\psi^{*}} &=& \delta_{\psi} I_{\psi}\big\vert_{\psi=\psi^{*}}
- \int \delta_{\psi}(\Psi[\psi] - \psi)\vert_{\psi=\psi^{*}} D[\psi^{*}] \nonumber\\
&& - \int(\Psi[\psi^{*}] - \psi^{*}) \delta_{\psi} D[\psi]\vert_{\psi=\psi^{*}} \nonumber\\
&=& 0.
\end{eqnarray}
The first term is zero from our previous deduction that $\psi^{*}$ extremizes the functional $I_{\psi}$. The second term vanishes by virtue of the definition of $\psi^{*}$. Finally, the third term equates to zero from Eq.~\eqref{eq:equalateqm}. And so we preserve property (\textbf{A}). Thus far, our conclusion is that the modified functional serves the same purpose as the original one. However, as we will show next, its second variation might well be different. 

We now compute the second variation of the modified reduced functional at equilibrium. This calculation begins by computing the second functional derivative of $F$, keeping in mind that at equilibrium $D[\psi^{*}] = 0$ and $\Psi[\psi^{*}] - \psi^{*} = 0$. 
We have from Eq.~\eqref{eq:fvar}:
\begin{eqnarray}\label{eq:svar1}
\delta^{2}_{\psi}F_{\psi}[\psi^{*}] &=& \delta^{2}_{\psi} I_{\psi}[\psi^{*}]
- \int \delta_{\psi(y)}(\Psi[\psi] - \psi)\big\vert_{\psi=\psi^{*}} \frac{\delta D[\psi^*]}{\delta\psi(x)} \nonumber \\
&&- \int\delta_{\psi(x)}(\Psi[\psi] - \psi)\big\vert_{\psi=\psi^{*}} \frac{\delta D[\psi^{*}]}{\delta\psi(y)}.
\end{eqnarray}
To continue with the evaluation we need to calculate the variation of the modified multiplier. We have from Eq.~\eqref{eq:modlm}
\begin{eqnarray}
\delta_{\psi(x)}\Psi[\psi]
&=&\delta_{u}h^{-1}(u)|_{u=D[\psi]+h(\psi)}\times\\\nonumber
&&\left(\delta_{\psi(x)} D[\psi(x')] + \delta_{\psi(x)} h(\psi(x')\right).
\end{eqnarray}
As we are interested in the properties of the second variation at equilibrium we note that $D[\psi^{*}]+h(\psi^{*})=h(\psi^{*})$, so that
\begin{eqnarray}
\delta_{\psi(x)}\Psi[\psi^{*}]
&=&\delta_{h}h^{-1}[h(\psi^{*}(x'))]\times\\
&&\left(\delta_{\psi(x)} D[\psi^{*}(x')]
+ \delta_{\psi(x)} h(\psi^{*}(x')\right)\nonumber.
\end{eqnarray}
We make use of the relation $h^{-1}(h(y))=y$, so that $\delta_{u=h(y)}h^{-1}(u)\delta_{y}h(y)=1$, to obtain:
\begin{eqnarray}\label{eq:varlm}
\delta_{\psi(x)}\Psi[\psi^{*}]&=&\delta_{h}h^{-1}[h(\psi^{*})]
\delta_{\psi(x)}D[\psi^{*}(x')] \nonumber \\
&&+ \delta_{\psi(x)}\psi(x')\vert_{\psi=\psi^{*}}.
\end{eqnarray}

Equation \eqref{eq:varlm} gives the variation of the modified multiplier at equilibrium. We obtain from this equation:
\begin{equation}\label{eq:diffvar}
\delta_{\psi(x)}(\Psi[\psi]-\psi(x'))\vert_{\psi=\psi^{*}}
=\delta_{h}h^{-1}[h(\psi^{*})]\delta_{\psi(x)}D[\psi^{*}(x')].
\end{equation}
We recognize that the term in the left-hand side of the above equation appears in Eq.~\eqref{eq:svar1}. Using Eq.~\eqref{eq:diffvar} to substitute for this term in Eq.~\eqref{eq:svar1}, and employing the latter in Eq.~\eqref{eq:secondvar} leads to the following expression for the second variation of $F_\psi$ at equilibrium:
\begin{eqnarray}\label{eq:svar2}
\delta^{2}F_{\psi}[\psi^{*}] &=& \delta^{2} I_{\psi}[\psi^{*}]\\\nonumber
&&- 2\int \delta_{h} h^{-1}[h(\psi^{*}(x'))](\delta D[\psi^{*}(x')])^{2}dx',
\end{eqnarray}
where $\delta D$ is given by Eq.~\eqref{eq:delD}.
As is clear from Eq.~\eqref{eq:svar2}, the introduction of the modified multiplier has augmented the second variation of the original functional $I_{\psi}$ by the second term on the left hand side. Finally, using Eq.~\eqref{eq:svarI}, which gives the result for the second variation of $I_{\psi}$, and adopting a compact notation, Eq.~\eqref{eq:svar2} becomes
\begin{equation}\label{eq:svar3}
\delta^{2}F_{\psi}[\psi^{*}] = -\int \delta D[\psi^{*}]\delta\psi
- 2\int \delta_{h} h^{-1}[h(\psi^{*})](\delta D[\psi^{*}])^{2}.
\end{equation}
Equation \eqref{eq:svar3} is the principal result of this work.
It gives the second variation of the modified reduced functional $F_{\psi}$ at equilibrium. We note that the sign of the second term on the right-hand side of Eq.~\eqref{eq:svar3} depends crucially on the properties of the function $h$. 
It is therefore possible, for at least some systems, to produce a functional with a positive-definite second variation by suitably choosing the function $h$. In this way, we can transform the original reduced functional into a functional that has property (\textbf{C}). We shall illustrate this below with many examples. We note that the negative sign preceding this term is not a large obstacle to render the expression positive-definite as we can always choose the function $h^{-1}$
to be monotonically decreasing. Finally, in terms of $F_{\rho}$ we can state the final result for the second variation as: 
\begin{eqnarray}
\delta^{2}F_{\rho}[\rho^{*}] &=&
- \int \delta S[\rho^{*}] \delta C[\rho^{*}] \\ \nonumber
&&- 2\int \delta_{h}h^{-1}[h(S[\rho^{*}])](\delta C[\rho^{*}])^{2},
\end{eqnarray}
where $\delta S[\rho]$ and $\delta C[\rho]$ are given by Eqs. \eqref{eq:delS} and \eqref{eq:delC} respectively.

It is useful to point out a particular case of our modified Lagrange multiplier method that leads to a simple procedure for transforming the constraint. Let us rewrite the reduced functional obtained via the standard Lagrange multiplier method:
\beq
I_{\psi}[\psi]=I_{o}[\psi]-\int\psi D[\psi].
\eeq
We choose $h(\psi)=-\psi$ to construct our modified multiplier obtaining
$\Psi[\psi]=\psi-D[\psi]$.
Carrying out the transformation, we see that it leads to the modified functional:
\beq
F_{\psi}[\psi]=I_{o}[\psi]-\int\psi D[\psi]+\int D[\psi]^{2}.
\eeq
It is clear that the modification does not disturb the equilibrium position since, upon variation, the contribution of the quadratic term in $D[\psi]$ vanishes at equilibrium and a solution of the original problem remains a solution. It is also clear why the procedure improves the behavior of the functional as it adds a clear positive-definite term at the equilibrium point. However, we note that some of the transformations we use in the examples that follow do not fit this simple template. 

\section{Examples}
The rather abstract nature of previous developments demands a variety of examples to clarify their meaning. We start with simple examples progressing to the most interesting ones. Certain aspects of the example of electrostatics in polarizable media and the example of Poisson-Boltzmann theory to describe charged systems
are treated in detail in previous publications \cite{jso2,jso3}, albeit without resort to the general theory presented here.

\subsection{The one-dimensional case}\label{sec:1d}
We begin our examples with a simple quadratic function before considering a functional \emph{per se}. In these first few cases, functionals are simply functions, and variations of functions become ordinary derivatives, and delta functions are replaced by Kronecker deltas. We consider the quadratic form
\beq
I_{o}(x)=\frac{1}{2}ax^{2}
\eeq
with $a > 0$.
Note that, considering the notations employed in the last few sections, we have $\rho=x$.
We impose the constraint that $x=c$, that is, we shall use
\beq
C(x)=x-c.
\eeq
Its obvious solution for $x$ is $x^{*}=c$, and we see that our function takes the value $I_{o}(x^{*})=ac^{2}/2.$ We can also note that $I_{o}$ is positive-definite everywhere. Clearly, this is a trivial problem with a unique solution, but it illustrates how the method works.

We set up the constrained minimization as
\beq
I(x;\psi)=\frac{1}{2}ax^{2}-\psi(x-c).
\eeq
Variation with respect to \textbf{$x$} leads to
\beq
Q(x,\psi)=ax-\psi=0.
\eeq
Clearly, this equation invites a simple change in variables, and we can define the $S$ and $R$ functions:
\beq\label{eq:1dS}
\psi = S(x) = ax
\eeq
and
\beq\label{eq:1dR}
x = R(\psi) = \psi/a.
\eeq

Let us first consider the reduced functional $I_{x}$, which is read from
Eq.~\eqref{eq:Irho} after carrying out the substitution using Eq.~\eqref{eq:1dS}.
We obtain the function
\beq
I_{x}(x)=-\frac{1}{2}ax^{2}+acx.
\eeq
The equilibrium equation is
\beq
-ax+ac=0.
\eeq
This is in agreement with our general theory as we obtain the above form from Eq.~\eqref{eq:eqmIrho}:
\beq
-C(x)\frac{dS(x)}{dx} =-(x-c)a =-ax+ac = 0.
\eeq
We emphasize here that the extremization of the reduced functional is equivalent to the solution of the constraint equation $C(x)=0$. Its solution is of course $x^{*}=c$, as in the original problem. To be more precise, we do require that the derivative $dS/dx$ (which is $a$) not be zero.

We obtain, for the second variation
\beq
\frac{d^{2}I_{x}}{dx^{2}}=-a,
\eeq
which, alternatively, can be read from the formula in Eq.~\eqref{eq:svarrho}:
\beq
-\frac{dC(x)}{dx}\frac{dS(x)}{dx}=-a.
\eeq
The price to be paid for the use of this formulation is that the second variation is negative. Our extremization is in fact a maximization.

We can take a look at the parallel reduced problem for $I_{\psi}$. The function to be extremized is:
\begin{equation}\label{eq:1dIpsi}
I_{\psi}(\psi)=-\frac{1}{2a}\psi^{2}+c\psi.
\end{equation}
The extremization of this functional leads to the equation
\beq
-\frac{1}{a}\psi+c=0.
\eeq
We can easily recognize in this expression the condition 
$-D(\psi)=-C(R(\psi))=-(R(\psi)-c)=-(\psi/a)+c=0$,
as developed in the general framework (see Eq.~\eqref{eq:eqmIpsi}). We recover the solution $\psi^{*}=ac$ so that $x^{*}=c$. We also note that the function $I_{\psi}$ is obviously negative-definite. Quickly, we can check that its second variation is:
\beq
\frac{d^{2}I_{\psi}}{d\psi^{2}}=-\frac{1}{a},
\eeq
or alternatively from Eq.~\eqref{eq:sfdI} we have
\beq
-\frac{dD(\psi)}{d\psi}=-\frac{1}{a},
\eeq
so that the two evaluations are equivalent. In this last expression the delta function in Eq.~\eqref{eq:sfdI} is replaced by a Kronecker delta on a single index (associated with the variable $\psi)$ which is simply 1.

We want to render our formulation of the reduced problem positive-definite.
To simply change the sign of the functional does not work since we would then change the evaluation at the minimum, and thus we would be violating one of the important properties, namely, property (\textbf{B}), of a useful functional. Also, in problems with multiple variables a global change in sign would modify
the properties of the functional with respect to all other variables. We thus proceed instead, with our general method. The modification framework is easier to follow in the $I_{\psi}$ format which we will use now.

We begin with choosing the function $h(\psi) = -\psi/a$. Then we can construct
the following modified constraint
\begin{eqnarray}
\Psi(\psi)&=&h^{-1}[D(\psi)+h(\psi)]\\\nonumber
&=&-a\left(\left(\frac{1}{a}\psi-c\right)-\frac{1}{a}\psi\right)=ac.
\end{eqnarray}
We have obtained a constant multiplier. The modified functional is
\begin{eqnarray}\label{eq:1dFpsi}
F_{\psi}(\psi)&=&I_{o}[R(\psi)]-\Psi(\psi)D(\psi)\\\nonumber
&=&\frac{1}{2a}\psi^{2}-ac\left(\frac{1}{a}\psi-c\right)
=\frac{1}{2a}(\psi-ac)^{2}+\frac{1}{2}ac^{2}.
\end{eqnarray}
Amazingly, this transformation has rendered the problem into a patently positive-definite one, which obviously has the expected minimum at $\psi^{*}=ac$ (or $x^{*}=c$), and maintains the correct evaluation of the functional. Let us confirm these properties of the above functional. The equilibrium equation is
\beq
\frac{1}{a}(\psi-ac)=0,
\eeq
which is identical to $D(\psi)=0$, as suggested by our general theory. We obtain for the solution $\psi^{*}=ac$, which is indeed the correct solution. At the equilibrium point, we can evaluate the second variation as follows:
\begin{eqnarray}
\frac{d^{2}F_{\psi}}{d\psi^{2}}\Big\vert_{\psi = \psi^{*}}
&=& -\frac{dD}{d\psi}\Big\vert_{\psi=\psi^{*}}\\\nonumber
&&-2\frac{dh^{-1}}{du}\Big\vert_{u=h(\psi)}\left(\frac{dD}{d\psi}\Big\vert_{\psi=\psi^{*}}\right)^{2}\\\nonumber
&=& -\frac{1}{a} - 2(-a)\left(\frac{1}{a}\right)^{2}
=\frac{1}{a}.
\end{eqnarray}
The result is positive-definite, confirming that our modified functional is convex.

\begin{figure}
\centerline{
\includegraphics[scale=0.72]{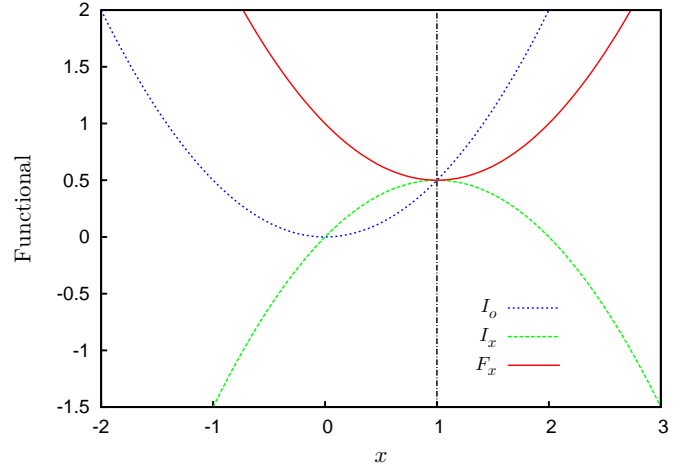}
}
\caption
{\label{fig1}
Functionals of base variable $x$ corresponding to the example of a one-dimensional quadratic form. The original (unconstrained) functional $I_{o}$ is represented by a dotted blue line. The reduced functional $I_{x}$ produced via the standard Lagrange multiplier method is the dashed green line. Solid red line is the functional $F_{x}$ that results from the application of the modified Lagrange multiplier procedure. The black dashed-dotted line is the constraint $C(x)$ for this example. The two functionals $I_{x}$ and $F_{x}$ meet at one point, the point of extremum, which lies on the constraint line. While $I_{x}$ is maximized at this extremum point, $F_{x}$ becomes a minimum. 
}
\end{figure}

We omit the full development of these results for the modified functional based on the variable $x$. We simply note that, using again $h(\psi)=-\psi/a$ and carrying out the substitution $\psi = S(x)$ in Eq.~\eqref{eq:1dFpsi}, we obtain the following functional for $x$:
\beq
F_{x}(x)=\frac{a}{2}(x-c)^{2}+\frac{1}{2}ac^{2}.
\eeq
This has the same basic properties as we deduced for the modified reduced functional for $\psi$, including positive-definiteness. In Fig.~\ref{fig1} we plot the functionals of variable $x$ obtained for this one-dimensional example. We employ $a = 1$ and $c = 1$, making $x=1$ as the obvious solution point and we expect the reduced functionals to be stationary at this particular point. It is clear from the figure that the standard Lagrange multiplier method produces a functional (dashed green line) that becomes a maximum at the extremum. On the other hand, the functional constructed via the modified Lagrange multiplier procedure (solid red line) becomes a minimum at the same extremum point. Figure \ref{fig1} provides a helpful visualization for the functionals obtained in the case of other, more complicated examples that follow, where the concave reduced functional is similarly transformed into a convex form.

Before leaving this example we point out that other choices of the function $h$ can be used. While we have made a perhaps ideal choice above, we have plenty of options. Consider for example the choice, in the spirit of the work we sketch below for the Poisson-Boltzmann equation, $h(\psi)=\exp(-\psi)$.
Employing this particular $h$ function, our reduced functional reads:
\beq
F_{\psi}(\psi)=\frac{1}{2a}\psi^{2}+\ln\left(\exp(-\psi)+\frac{1}{a}\psi-c\right)\left(\frac{1}{a}\psi-c\right).
\eeq
We find that we still have a solution at $\psi=ac$, but as the reader would note, the nonlinearity of the expression reduces the region of definition of the function. Computing the second derivative of the function we obtain:
\begin{eqnarray}
\frac{d^{2}F_{\psi}}{d\psi^{2}}&=&
\frac{1}{a}+\frac{2}{a}\frac{1-a\exp(-\psi)}{(a\exp(-\psi)+\psi-ac)}
\\\nonumber
&&+\frac{1}{a}\frac{a\exp(-\psi)(\psi+2-ac)-1}{(a\exp(-\psi)+\psi-ac)^{2}}(\psi-ac), 
\end{eqnarray}
which reduces at equilibrium to
\beq\label{eq:svar_exp}
\frac{d^{2}F_{\psi}}{d\psi^{2}}\Big\vert_{\psi=\psi^{*}}=\frac{1}{a}+2\frac{1-a\exp(-ac)}{a^{2}\exp(-ac)}.
\eeq
For $c>0$, if our coefficient $a$ satisfies $0<a<1$, this second derivative is positive-definite at and near its extremum, which is therefore a minimum.
This last expression could also be read from our general results:
\begin{eqnarray}
\frac{d^{2}F_{\psi}}{d\psi^{2}}\Big\vert_{\psi=\psi^{*}}&=&
-\frac{dD}{d\psi}\Big\vert_{\psi=\psi^{*}} \\\nonumber
&&- 2\frac{dh^{-1}}{du}\Big\vert_{u=h(\psi)}\left(\frac{dD}{d\psi}\Big\vert_{\psi=\psi^{*}}\right)^{2}\\\nonumber
&=&2\frac{1}{a^{2}\exp(-ac)}-\frac{1}{a}
\end{eqnarray}
which is the same result as Eq.~\eqref{eq:svar_exp}.

\subsection{A vector space}
We consider now a less trivial example, using a quadratic function in a multi-dimensional space and multiple linear constraints. A key difference with the previous case is the loss of a one-to-one relationship between the variable $x$
and the multiplier $\psi$. We consider the quadratic form
\beq
I_{o}=\frac{1}{2}x_i A_{ij} x_j,
\eeq
where $x_i$ is a vector in a real Euclidean $N-$dimensional space. For simplicity, we take the matrix $A_{ij}$ as positive-definite and we take it to be symmetric. We use the repeated index summation convention. We impose the constraint that the vectors should live in the intersection of $M$ hyperplanes:
\beq
C^{\alpha}(x_i) = n^{\alpha}_i x_i-c^{\alpha} = 0.
\eeq
Here, for each constraint index $\alpha$, we have a vector  $n^\alpha_i$ that can be considered normal to the constraint hyperplane. Note that the superscript ranges over indices from $1$ to $M$, so that, considered as a matrix, $n^\alpha_i$ has $M\times N$ elements. We employ the convention where the superscript (Greek letters) specifies the constraint index and subscript denotes the base vector ($x$) component. Note that $c^{\alpha}$ is a matrix containing $M$ constant elements.

We set up the function
\beq
I(x_i;\psi^\alpha)=\frac{1}{2}x_iA_{ij}x_j-\psi^{\alpha}(n^{\alpha}_i x_i - c^{\alpha}).
\eeq
Our multiplier $\psi^\alpha$ is a $M-$dimensional vector. Variation with respect to $x$ leads to
\beq
A_{ij}x_j - n^\alpha_i \psi^\alpha = 0,
\eeq
where we limit the space of variables only to those values of $x_i$ where the above equation can be solved, namely, the linear subspace of vectors of the form $x_i=A^{-1}_{ij}n^\alpha_j\psi^\alpha$. Letting the constraint variable range over all its possible values, we note that this subspace is only $M-$dimensional.

To explicitly write the relations between the coordinate vector $x_i$ and the multiplier $\psi^\alpha$, we first introduce the auxiliary square matrices $P^{\alpha\beta}=n^{\alpha}_i n^\beta_i$ and $V^{\alpha\beta}=n^\alpha_i A^{-1}_{ij} n^\beta_j$. Their inverses $(P^{-1})^{\alpha\beta}$, $(V^{-1})^{\alpha\beta}$, are defined in the obvious way. Note that the matrices $P$ and $V$ are both of dimensions $M\times M$ and the repeated index in their definition implies the sum over the $N$ components corresponding to the base vector $x$. When we restrict ourselves to the valid space of variables as noted above, we have a map between the values of the two sets of variables $x_i$ and $\psi^\alpha$:
\beq
\psi^\alpha = S(x_i)= (P^{-1})^{\alpha\beta}n^{\beta}_i A_{ij} x_j,
\eeq
with the inverse relation:
\beq
x_i = R(\psi^\alpha)= A^{-1}_{ij}n^\alpha_j\psi^\alpha.
\eeq
We can now obtain the functions
\beq
D^\alpha(\psi^\beta) = C^\alpha(R(\psi^\beta)) =  V^{\alpha\beta}\psi^\beta - c^\alpha.
\eeq
These functions produce vectors, and we emphasize this fact by adding relevant subindices to their symbols. The condition $D^\alpha(\psi^\beta)=0$ is in fact an ensemble of $M$ conditions. Now, the reduced functional $I_{\psi}$, from  Eq.~\eqref{eq:Ipsi}, reads
\beq
I_{\psi}(\psi^\alpha)=-\frac{1}{2}\psi^\alpha V^{\alpha\beta} \psi^\beta + \psi^\alpha c^\alpha,
\eeq
which can be shown to be negative-definite.
Variation of the above functional leads to a result that, when restricted to the proper subspace, is identical to:
\beq
n^\alpha_i x_i - c^\alpha = 0.
\eeq
Namely, our result is consistent with the claim that the equilibrium equation is simply the overall constraint itself.

We can also check the validity of our expressions for the second variation of the functional. We have
\beq
\frac{\delta^{2}I_{\psi}}{\delta\psi^\alpha \delta\psi^\beta} = -V^{\alpha\beta},
\eeq
which indeed has the form of $-\delta D^\alpha(\psi^\beta)/\delta\psi^\beta$.

As in the one-dimensional case, we can \emph{fix} our functional by using the function $h(\psi^\alpha) = -V^{\alpha\beta} \psi^\beta$. This results in a modified multiplier
\beq
\Psi^\alpha=c^\gamma (V^{-1})^{\gamma\alpha}
\eeq
and a final result for the functional similar to the one-dimensional case:
\beq
F_{\psi}(\mathbf{\psi^\alpha})=\frac{1}{2} \psi^\alpha V^{\alpha\beta} \psi^\beta
- c^\gamma (V^{-1})^{\gamma\alpha} V^{\alpha\beta} \psi^\beta + c^\gamma (V^{-1})^{\gamma\alpha} c^\alpha. 
\eeq
The above functional is positive-definite.

In this example, it becomes clear that our work with the reduced functional takes place, in general, in a smaller space than the original region of the definition of the variable $x$. The more interesting examples below show a similar situation within the context of infinite-dimensional function spaces. 

\subsection{Electrostatics in free space}\label{sec:efs}
We now proceed to more complex and practical examples. We shall produce many different presentations of the electrostatic variational problem. Gaussian units will be adopted in what follows.

We start, in all cases, with the following expression for the energy of the electric field $\mathbf{E}$:
\beq
I_{o}[\mathbf{E}]=\frac{1}{8\pi}\int\mathbf{E}\cdot\mathbf{E}.
\eeq
The minimum of this expression is zero. To create non-trivial fields, there must be charges and Gauss's law must be obeyed,
\beq
C[\mathbf{E}]=\frac{1}{4\pi}\mathbf{\nabla}\cdot\mathbf{E} - \rho = 0,
\eeq
which we take as a constraint on the form of the field. Here, $\rho$ denotes the charge density which is considered as a parameter field, not subjected to 
variations. We obtain the standard form of the constrained functional similar to Eq.~\eqref{eq:CF1}:
\beq
I[\mathbf{E};\phi]=\frac{1}{8\pi}\int\mathbf{E}\cdot\mathbf{E}-
\int\phi\left(\frac{1}{4\pi}\mathbf{\nabla} \cdot \mathbf{E}-\rho\right)
\eeq
where $\phi$ is the Lagrange multiplier and it will be soon identified with the electric potential. 

Variation with respect to the field and integration by parts gives us the relation
\beq
Q(\mathbf{E};\phi)=\frac{1}{4\pi}(\mathbf{E}+\mathbf{\nabla}\phi) = 0.
\eeq
Compared to the previous examples, we now have a rather non-trivial functional connecting the original field and the multiplier:
\beq
\mathbf{E} = R[\phi] = \mathbf{-\nabla}\phi.
\eeq
We can also obtain the multiplier as a function of the original field:
\begin{equation}\label{eq:efs_S}
\phi(\mathbf{x}) = S[\mathbf{E}] = 
\frac{1}{4\pi}\int G(\mathbf{x},\mathbf{y})\nabla\cdot\mathbf{E}(\mathbf{y})
\end{equation}
where $G(\mathbf{x},\mathbf{y})$ is the electrostatic Green's function given as
\beq
G(\mathbf{x},\mathbf{y})=\frac{1}{|\mathbf{x-y}|}.
\eeq
To simplify the notation, we will write 
$\int d\mathbf{y\:}G(\mathbf{x},\mathbf{\mathbf{y}})h(\mathbf{y)}$
as $\int Gh$ when it is unlikely to cause confusion.

Not every vector field can be expressed as a gradient, but since the equation $Q(\mathbf{E},\phi)=0$ is only valid for these cases, we can restrict the fields considered to those admitting a potential. We also note that in writing the functional $S[\mathbf{E}]$ in Eq.~\eqref{eq:efs_S}, we have used boundary conditions that eliminate constant fields that do not vanish at infinity. 

From Eq.~\eqref{eq:Ipsi}, we obtain the following expression for the reduced functional:
\begin{eqnarray}\label{eq:efs_Iphi}
I_{\phi}[\phi]&=&\frac{1}{8\pi}\int\mathbf{\nabla}\phi\cdot\mathbf{\nabla}\phi
-\int\phi\left(-\frac{1}{4\pi}\nabla^{2}\phi - \rho\right)\\\nonumber
&=&-\frac{1}{8\pi}\int\mathbf{\nabla}\phi\cdot\mathbf{\nabla}\phi+\int\phi\rho,
\end{eqnarray}
where the second equality of the above equation follows from integration by parts.
The above functional is well known \cite{schwinger} and one can tell by inspection that it is 
negative-definite. We can use the formula for the second variation of a standard reduced functional derived in Sec.~\ref{sec:stdvar} to confirm our suspicion. From the first equality of Eq.~\eqref{eq:efs_Iphi}, we note the form of the constraint, 
\beq\label{eq:efs_D}
D[\phi]=-\frac{1}{4\pi}\nabla^{2}\phi-\rho.
\eeq
Quickly, we compute the variation $\delta D$ defined in Eq.~\eqref{eq:delD}:
\beq\label{eq:efs_delD}
\delta D = - \frac{1}{4\pi} \nabla^{2} \delta \phi.
\eeq
Finally, using the above result in Eq.~\eqref{eq:svarIpsi} we obtain the second variation of $I_{\phi}$ at equilibrium to be
\beq
\delta^{2}I_{\phi}[\phi^{*}] = - \int \delta D \delta \phi = 
\frac{1}{4\pi}\int \delta \phi \nabla^{2} \delta \phi = 
- \frac{1}{4\pi}\int |\nabla\delta\phi|^{2} .
\eeq
Clearly, the above second variation is strictly negative. 

We now use the modified multiplier procedure to obtain a convex functional for the electrostatics of charges in free space. In Eq.~\eqref{eq:modlm}, using the functional 
$h(\phi)=(4\pi)^{-1}\nabla^{2}\phi$,
so that $h^{-1}(\sigma)=-\int G\sigma$, we obtain the following form for the 
modified Lagrange multiplier $\Phi$:
\beq\label{eq:efs_h-1}
\Phi = h^{-1}[D[\phi]+h(\phi)]=h^{-1}(-\rho)=\int G\rho.
\eeq
Note that we have obtained a multiplier that does not depend on $\phi$. Using this result in Eq.~\eqref{eq:Fpsi} gives the modified reduced functional:
\begin{align}\label{eq:efs_Fpsi}
F_{\phi}[\phi]&=\frac{1}{8\pi}\int\mathbf{\nabla}\phi\cdot\mathbf{\nabla}\phi
-\iint G\rho\left(-\frac{1}{4\pi}\nabla^{2}\phi - \rho\right)\notag\\
&=\frac{1}{8\pi}\int\mathbf{\nabla}\phi\cdot\mathbf{\nabla}\phi-\int\phi\rho+\iint\rho G\rho.
\end{align}
Judging by the form of the second equality above, it is easy to tell that we have rendered the functional positive-definite. Once again, we can confirm this by using Eq.~\eqref{eq:svar3}, the formula for the second variation of the modified reduced functional at equilibrium. In this regard, it is useful to know that for the present system, with our particular choice of $h(\phi) = (4\pi)^{-1}\nabla^{2}\phi$, we have 
\beq
\frac{\delta h(\phi)}{\delta\phi} =
-\frac{\delta D}{\delta\phi}.
\eeq 
Remembering once again that 
\beq
\frac{\delta h^{-1}[u]}{\delta u}\Big\vert_{u = h} = 
\left(\frac{\delta h (v)}{\delta v}\Big\vert_{v = h^{-1}}\right)^{-1}
\eeq
and, at equilibrium, $\Phi = h^{-1}[h(\phi)] = \phi$, 
we obtain from Eq.~\eqref{eq:svar3}, 
\begin{align}\label{eq:efs_svarF}
\delta^{2}F_{\phi}[\phi^{*}] &= 
- \frac{1}{4\pi}\int |\nabla\delta\phi|^{2} 
+ 2\iint \delta(x - x')\delta\phi(x) \,\delta D(x') \notag\\
&=  - \frac{1}{4\pi} \int |\nabla\delta\phi|^{2} + 2 \frac{1}{4\pi} \int |\nabla\delta\phi|^{2} \notag\\
&=  \frac{1}{4\pi} \int |\nabla\delta\phi|^{2}. 
\end{align}
In the above set of equations, the second equality follows from using Eq.~\eqref{eq:efs_delD} and carrying out the integral over $x$. Clearly, from the final equality in Eq.~\eqref{eq:efs_svarF}, we conclude that the second variation of $F_{\phi}$ is strictly positive.

It is useful to take a closer look at $F_{\phi}$, in particular, the way it is expressed in the second equality of Eq.~\eqref{eq:efs_Fpsi}. This result is deceptively simple. It might appear that we have simply reversed the signs of the original result in Eq.~\eqref{eq:efs_Iphi} and ``fixed'' the value of the energy by adding a term. However, as derived from our general framework, this is simply one of infinite equivalent functionals that indeed recovers the original equations of equilibrium and evaluates to the correct value of the energy while preserving the minimum property. We can present this functional in yet another
rather suggestive way. Let us change variables replacing the potential for an equivalent distribution of charge $\sigma,$  so that $\nabla^{2}\phi=-4\pi\sigma$.
Imposing suitable boundary conditions and restricting the space of functions allowed for both the variables $\phi$ and $\sigma$, we can make this a one to one function. While the construction of the inverse is equivalent to the solution of the constraint, conceptually these two operations are different; we change the variational functional variable but we are not solving for the equilibrium. We obtain, after some manipulations:
\beq\label{eq:efs_Fsigma}
F_{\sigma}[\sigma]=\frac{1}{2}\iint\sigma G(\sigma-2\rho)
+\iint\rho G\rho.
\eeq
This simple form can be easily interpreted. To an arbitrary potential field
configuration we can associate a fictitious charge density $\sigma$ by taking the Laplacian of the potential. The total energy of this field, is composed of its self interaction and the interaction with the actual charge $\rho$. 
The energy achieves its minimum when the fictitious charge equals the actual charge. This seemingly innocuous statement is actually the basis of very useful methods to analyze more complicated systems, such as the case of particles in a polarizable medium which is our next example.

Finally, we note that the standard Lagrange multiplier procedure evaluates to the
following functional in the variable $\mathbf{E}$:
\beq
I_{E}[\mathbf{E}] = -\frac{1}{8\pi}\int\mathbf{E}\cdot\mathbf{E} 
+\iint \rho G\frac{1}{4\pi}\nabla\cdot\mathbf{E},
\eeq
which is a negative-definite form. Once again, substituting $\phi$ from Eq.~\eqref{eq:efs_S}
in terms of $\mathbf{E}$ in Eq.~\eqref{eq:efs_Fpsi}, generates the modified 
reduced functional of variable $\mathbf{E}$,
\beq
F_{E}[\mathbf{E}] = 
\frac{1}{8\pi}\int \mathbf{E}\cdot\mathbf{E} 
- \iint \rho G \frac{1}{4\pi}\nabla\cdot\mathbf{E} + \iint\rho G \rho,
\eeq
which has the property of being positive-definite at its extremum.

\subsection{Electrostatics in polarizable medium}\label{sec:electrostatics_media}
The motivation for the investigations carried out in this subsection is to construct a positive-definite functional for simulating charges in a dielectric medium with spatially varying permittivity. We have presented the basic results in our previous publications \cite{jso1,jso2}. Here, we wish to show how those results fit within the modified functional method. 

The starting point is the energy, expressed in terms of the electric field $\mathbf{E}$ and the polarization vector $\mathbf{P}$. The net charge is composed of the free charge density $\rho$ and the induced polarization charge density $\omega=-\mathbf{\nabla}\cdot\mathbf{P}$.
Our initial functional is then 
\begin{eqnarray}
I[\mathbf{E},\mathbf{P};\phi]&=&\frac{1}{8\pi}\int\mathbf{E\cdot E}
+\int\frac{1}{2\chi}\mathbf{P}\cdot\mathbf{P}\nonumber\\
&&-\int\phi\left(\frac{1}{4\pi}\mathbf{\nabla}\cdot\mathbf{E}-\mathbf{\nabla}\cdot\mathbf{P} - \rho\right)
\end{eqnarray}
where, once again, we include Gauss's law as a constraint to the electrostatic energy. The functional has base variables $\mathbf{E}$ and $\mathbf{P}$ and
multiplier $\phi$. A first reduction of the functional by elimination of the electric field and potential leads to the functional
\begin{eqnarray}\label{eq:em_fnalP}
I_{P}[\mathbf{P}]&=&\frac{1}{8\pi}\int\frac{\epsilon}{\chi^{2}}\mathbf{P}\cdot\mathbf{P}\\\nonumber
&&-\iint G\left(\rho-\mathbf{\nabla}\cdot\mathbf{P}\right)\left(\mathbf{\nabla}\cdot\frac{\epsilon}{4\pi\chi}\mathbf{P}-\rho\right).
\end{eqnarray}
This presentation of the functional still exhibits the structure of our formal derivations with 
\beq\label{eq:constraint_CP}
C[\mathbf{P}]=\mathbf{\nabla}\cdot\frac{\epsilon}{4\pi\chi}\mathbf{P} - \rho = 0,
\eeq
and 
\beq
\psi[\mathbf{P}]=\int G(\rho - \mathbf{\nabla}\cdot\mathbf{P}).
\eeq
We have for the extremum condition of $I_{P}[\mathbf{P}]$:
\begin{equation}\label{eq:em_extcond}
\frac{1}{\chi}\mathbf{P} + \mathbf{\nabla} \int G(\rho- \mathbf{\nabla}\cdot\mathbf{P})=0,
\end{equation}
and formal manipulations reduce it to the previous form of the constraint $C[\mathbf{P}] = 0$ as given by Eq.~\eqref{eq:constraint_CP}. As expected, the equilibrium condition is equivalent to the constraint.

Equation \eqref{eq:em_extcond} has a clear interpretation. The integral gives the value of the potential created by the free and polarization charges, and the gradient of the potential reproduces the electric field which cancels the equivalent term $\mathbf{P}/\chi$. Rearrangement of the functional in Eq.~\eqref{eq:em_fnalP} produces
\beq
I_{P}[\mathbf{P}]=\int\frac{1}{2\chi}\mathbf{P}\cdot\mathbf{P}
+\frac{1}{2}\iint(\rho-\mathbf{\nabla}\cdot\mathbf{P})G(\rho-\mathbf{\nabla}\cdot\mathbf{P}).
\eeq
This functional, obtained elsewhere \cite{marcus,felderhof}, is clearly 
positive-definite and has been employed in simulation methods which rely on exploiting this property \cite{marchi}. A significant drawback of this expression is that it relies on a vector variable. 

To further take advantage of the functional in Eq.~\eqref{eq:em_fnalP}, we seek to replace the base variable $\mathbf{P}$ with scalar variables such as the potential $\phi$ or the scalar polarization density $\omega$. In particular, the transition to $\omega$ as the base variable offers numerous benefits when simulating charges in piecewise-uniform dielectric media as argued elsewhere \cite{allen1,jso1,boda}. The change of variables is arduous since the polarization vector is not solely determined by a potential. It can be checked, however, that the following expression can be used:
\beq
\mathbf{P}= - \chi\mathbf{\nabla}\int G(\rho+\omega).
\eeq
As before, we must emphasize that, in spite of its appearance, this is not a solution of any of the equations of the system, but simply a change in variables suggested by the known physical properties of the system. We can thus write:
\begin{eqnarray}
I_{\omega}[\omega]
&=& \iiint\frac{\chi}{2}\mathbf{\nabla} G(\rho+\omega)\cdot\nabla G(\rho+\omega)\\\nonumber
&&+\frac{1}{2}\int\left(\rho - \mathbf{\nabla}\cdot\chi\mathbf{\nabla}\int G(\rho+\omega)\right)
G \\\nonumber
&&\,\,\,\,\,\quad\quad\times\left(\rho - \mathbf{\nabla}\cdot\chi\mathbf{\nabla}\int G(\rho+\omega)\right).
\end{eqnarray}
This is again a positive-definite functional of the polarization charge $\omega$. 
We can simplify to:
\begin{equation}\label{eq:em_Iw}
I_{\omega}[\omega]
=\frac{1}{2}\iint\rho G (\rho+\Omega)
-\frac{1}{2}\iint\Omega G (\omega - \Omega)
\end{equation}
where we use the shorthand
\begin{equation}\label{eq:Omega}
\Omega=\mathbf{\nabla}\cdot\chi\mathbf{\nabla}\int G(\rho+\omega).
\end{equation}

Equation \eqref{eq:Omega} is clearly a recasting of the calculation of the induced polarization charge given the values of density of free charges and the
polarization charges. That is, the expression can be understood as a recursive calculation of the polarization charge. At equilibrium we must have $\Omega=\omega$ and in fact this requirement is the extremum condition that results from the variation of $I_{\omega}[\omega]$ with respect to $\omega$. 

In a previous work \cite{jso1}, we derived the functional in Eq.~\eqref{eq:em_Iw} in an indirect way, implementing the change of variables through a Lagrange multiplier. This is equivalent to the direct computation sketched above. However, using the formalism of modified multipliers presented here, different functionals can be obtained. We begin, like in Ref.~\cite{jso1}, by implementing the change of variable in the following way:
\begin{eqnarray}
I[\mathbf{P},\omega] &=& \int\frac{|\mathbf{P}|^{2}}{2\chi}
+ \frac{1}{2}\iint\left( \rho + \omega \right) G
\left(\rho + \omega \right)\nonumber \\
&&-\int \phi\left(\omega + \mathbf{\nabla}\cdot\mathbf{P}\right),
\end{eqnarray}
where $\phi$ is the Lagrange multiplier that will turn out to be the electrostatic potential. The above functional has two base variable fields
$\mathbf{P}$ and $\omega$ and the formalism of obtaining standard reduced functional in terms of $\phi$ as developed in Sec.~\ref{sec:stdvar} still goes through, producing first the $R$ functions that relate $\mathbf{P}$ and $\omega$ to $\phi$:
\beq\label{eq:P_R}
\mathbf{P} = R_{1}[\phi] = - \chi \mathbf{\nabla}\phi,
\eeq
\beq\label{eq:omega_R}
\omega = R_{2}[\phi] = - \frac{1}{4\pi}\nabla^{2}\phi - \rho.
\eeq
Using these functions, the standard procedure leads to the reduced functional
\begin{eqnarray}
I_{\phi}[\phi] &=& 
\int \frac{\chi}{2} \mathbf{\nabla}\phi \cdot \mathbf{\nabla}\phi
+ \frac{1}{2} \int \nabla^{2}\frac{\phi}{4\pi}  G \nabla^{2}\frac{\phi}{4\pi} \nonumber\\
&&- \int \phi \left(-\frac{1}{4\pi}\nabla^{2}\phi - \rho - \nabla\cdot\chi\nabla\phi \right),
\end{eqnarray}
which after some simple manipulations can be rendered to the following final form:
\beq\label{eq:em_Iphi}
I_{\phi}[\phi] = -\int \frac{\epsilon}{8\pi} \mathbf{\nabla}\phi \cdot \mathbf{\nabla}\phi
+ \int \phi\rho.
\eeq
It is clear that this functional is negative-definite. In fact, for the case of 
free space everywhere ($\epsilon = 1$), the above functional is the same as the functional in Eq.~\eqref{eq:efs_Iphi}. Once again, we can render it positive-definite by appropriately choosing the modified Lagrange multiplier.
We choose $h(\phi) = (4\pi)^{-1}\nabla^{2}\phi$ with its inverse given by 
$h^{-1}(y) = - \int G y$. Following Eq.~\eqref{eq:Fpsi}, we obtain the modified
reduced functional:
\begin{eqnarray}\label{eq:em_Fphi}
F_{\phi}[\phi] &=& \int \frac{\epsilon}{8\pi}\mathbf{\nabla}\phi\cdot\mathbf{\nabla}\phi \\\nonumber
&&+ \iint G\left(\mathbf{\nabla}\cdot\chi\mathbf{\nabla}\phi + \rho \right)
\left( \mathbf{\nabla}\cdot\frac{\epsilon\mathbf{\nabla}\phi}{4\pi} + \rho \right).
\end{eqnarray}
It can be checked from Eq.~\eqref{eq:svar3} that the above functional produces a 
positive-definite second variation at its extremum. We note that for the 
aforesaid case of $\epsilon = 1$ (equivalently, $\chi = 0$), the above functional is identical to the functional in Eq.~\eqref{eq:efs_Fpsi}.

The reduced functional can also be obtained with $\omega$ as the sole base variable. We first compute the inverse function from Eq.~\eqref{eq:omega_R}:
\beq
\phi = S_{2}[\omega] = \int G (\rho + \omega).
\eeq
Next, we substitute $\phi$ in terms of $\omega$ in Eq.~\eqref{eq:em_Fphi}, 
giving, after a few manipulations, the functional
\begin{equation}\label{eq:em_Fw}
F_{\omega}[\omega] =
\frac{1}{2}\iint \rho G \left(\rho + \omega\right)
- \iint \frac{1}{2}G (\rho - \omega + 2\Omega) \left(\omega - \Omega\right).
\end{equation}
One can check that, at equilibrium, the above functional exhibits positive second 
variation. Also, this functional is different from the functional $I_{\omega}[\omega]$ of Eq.~\eqref{eq:em_Iw}, which was derived via a direct change of variables. One can see the differences more clearly after rearranging the functional in Eq.~\eqref{eq:em_Iw} as:
\beq\label{eq:em_Iw2}
I_{\omega}[\omega] = 
\frac{1}{2}\iint \rho G \left(\rho + \omega\right) - 
\iint \frac{1}{2}G \left(\rho+\Omega\right) 
\left(\omega - \Omega\right).
\eeq
Clearly, both functionals $F_{\omega}$ and $I_{\omega}$ exhibit the structure of our formal derivations with the first terms in Eqs.~\eqref{eq:em_Fw} and \eqref{eq:em_Iw2} corresponding to the electrostatic energy and the second term harboring the constraint surface that $\omega$ must live on, namely, $\omega - \Omega = 0$. Moreover, we can now promptly identify that these two functionals differ only in the choice of the Lagrange multiplier that constraints the recursive relation $\omega - \Omega = 0$. 

While both $F_{\omega}$ and $I_{\omega}$ are useful functionals for analytical or computational investigations of the problem of charges in heterogeneous dielectric media, it is worth noting that $F_{\omega}$ offers the additional advantage of being useful for the case of charges in uniform vacuum. For this simple case, the functional $I_{\omega}$ identically equates to $\frac{1}{2}\iint \rho F \rho$, thus becoming independent of the base functional variable $\omega$ 
and consequently unusable. On the other hand, it is easy to see from Eq.~\eqref{eq:em_Fw} that for uniform vacuum
\beq
F_{\omega}[\omega;\chi=0] = \frac{1}{2}\iint \rho G \rho + \frac{1}{2}\iint \omega G \omega.
\eeq
Clearly, setting the first variation of the above functional to zero leads to the relation $\omega = 0$, which is indeed the correct solution for the case of free space. It is also important to note the contrast between $F_{\omega}[\omega;\chi=0]$ and $F_{\sigma}$, the functional of the charge density $\sigma$ produced in Eq.~\eqref{eq:efs_Fsigma}, remembering that $\omega$ is an actual charge density while $\sigma$ is a fictitious one.

In the end, we want to emphasize that in recent literature, the route to render the standard electrostatics functionals positive-definite has involved the cost of employing vectors such as $\mathbf{D}$ or $\mathbf{E}$ as base variables for the functional \cite{pujos}. In contrast, here we produce convex functionals for electrostatics which employ scalar functions such as $\phi$ or $\omega$ as base variables. Therefore, additional numerical costs due to the use of vector variables can be entirely avoided and in this light we expect our functionals to be excellent candidates for numerical minimization methods associated with 
simulations of charged systems.

\subsection{Poisson-Boltzmann theory for one-component plasma}
In a previous publication \cite{jso3} we have provided a detailed treatment of the variational framework for the Poisson-Boltzmann theory. Here, we simply sketch the role of the modified multiplier in producing positive-definite versions of the variational principle for the system. 

We consider the slightly simpler case of one-component plasma, which is interesting in its own right. This system has only one species of mobile ions with concentration $c$ and charge $q$. The condition of electroneutrality implies that there must be a compensating background charge of opposite sign and we denote its (charge) density as $\rho_{f}$. For simplicity, we will set $\epsilon=1$ everywhere. Our starting functional then simply adds to our electrostatic functional, terms that provide the Boltzmann relation between potential and local charge density:
\beq\label{eq:pb_Io}
I_{o}[\mathbf{E},c] =
\frac{1}{8\pi}\int\left|\mathbf{E}\right|^{2}
+\frac{1}{\beta}\int\left(c\,\textrm{ln}\left(c\Lambda^{3}\right) - c\right)
- \int \mu c,
\eeq
where $\Lambda$ and $\mu$ are, respectively, the deBroglie wavelength and chemical potential associated with the mobile ions and $\beta$ is the inverse thermal energy. 
Once again, we include Gauss's law as a constraint to the above functional 
obtaining
\beq
I[\mathbf{E},c,\phi] = I_{o}[\mathbf{E},c] - \int \phi 
\left(\frac{1}{4\pi}\mathbf{\nabla}\cdot\mathbf{E} - \rho_{f} - qc\right),
\eeq
where $\phi$ is the associated Lagrange multiplier and will soon turn out to be electrostatic potential. Variations of the above functional with respect to $\mathbf{E}$ and $\rho$ and subsequent substitution in favor of $\phi$ leads to the reduced functional
\begin{eqnarray}\label{eq:stdpb}
I_{\phi}[\phi]&=&
\int\left( \frac{1}{8\pi}\left|\nabla\phi\right|^{2}
-\frac{1}{\beta}C e^{-\beta q\phi}\left(\beta q\phi + 1\right)\right)\\\nonumber
&& - \int\phi
\left(-\frac{1}{4\pi}\mathbf{\nabla}\cdot\mathbf{\nabla}\phi 
- \rho_{f} - C q e^{-\beta q\phi} \right),
\end{eqnarray}
where $C$ is a constant made up from a combination of terms containing $\Lambda$ and $\mu$. $C$ can be interpreted as the bulk concentration. The last term of the above functional can be recognized as the product of the original multiplier times the constraint
\beq
D[\phi]=-\frac{1}{4\pi}\mathbf{\nabla}\cdot\mathbf{\nabla}\phi 
- \rho_{f} - C q e^{-\beta q\phi}.
\eeq
We can use formulas derived in Sec.~\ref{sec:stdvar} to evaluate the second variation of the above functional. We promptly compute the variation $\delta D$ given by Eq.~\eqref{eq:delD} for the above constraint as
\beq\label{eq:pb_delD}
\delta D = -\frac{1}{4\pi} \mathbf{\nabla}\cdot\mathbf{\nabla} \delta \phi
+ \beta C q^{2} e^{-\beta q \phi} \delta\phi,
\eeq
and using the result in Eq.~\eqref{eq:svarIpsi}, we obtain the second variation at equilibrium to be 
\beq\label{eq:pb_svarI}
\delta^{2} I_{\phi}[\phi^{*}]
= -\frac{1}{4\pi}\int |\nabla \delta\phi|^{2} 
- \beta C q^{2} \int e^{-\beta q \phi^{*}} (\delta\phi)^{2}.
\eeq
Clearly, the above second variation is strictly negative, implying that the reduced functional $I_{\phi}$ is negative-definite.

We now use the modified Lagrange multiplier method to obtain a convex functional for this problem. First, we choose the $h$ function to be $h(\phi)=Cq\exp(-\beta q \phi)$.
This form is suggested by the need of an appropriate amount of  negative contribution from the derivative of $h$ as seen in Eq.~\eqref{eq:svar3}, and the eventual tractability and simplicity of the form of the modified multiplier $\Phi$. We quickly find the inverse function to be $h^{-1}(y) = - (\beta q )^{-1} \textrm{ln} (y/C q)$. Using these functions we can transform the multiplier as prescribed in Eq.~\eqref{eq:modlm} and obtain
\beq
\Phi[\phi]=-\frac{1}{\beta q }\textrm{ln}
\left[\frac{-\nabla^{2}\phi - \rho_{f} }{Cq}\right].
\eeq
As in the example of Sec.\ref{sec:1d}, application of this modified multiplier demands restrictions on the range of the function space explored by $\phi$ so as to avoid, for example, imaginary values for the logarithm. 
The modified functional is
\begin{eqnarray}\label{eq:modpb}
F_{\phi}[\phi]&=&
\int\left( \frac{1}{8\pi}\left|\nabla\phi\right|^{2}
-\frac{1}{\beta}C e^{-\beta q\phi}\left(\beta q\phi + 1\right)\right)\\\nonumber
&& - \frac{-1}{\beta q }\int \textrm{ln}
\left[\frac{-\nabla^{2}\phi - \rho_{f} }{Cq}\right]\\\nonumber
&& \qquad\qquad\,\,\,\times \left(-\frac{1}{4\pi} \nabla^{2} \phi 
- \rho_{f} - C q e^{-\beta q\phi} \right).
\end{eqnarray}

We now calculate, from Eq.~\eqref{eq:svar3}, the second variation of the modified functional. We already have the expression for $\delta D$ given by Eq.~\eqref{eq:pb_delD}. We need $\delta_{h}h^{-1}(h)$, which is 
\beq\label{eq:pb_dh-1}
\delta_{h}h^{-1}(h(\phi^{*}))= \frac{1}{\delta_{\phi}h(\phi)}
=-\frac{e^{\beta q \phi^{*}}}{\beta C q^{2}}.
\eeq
The expression for the second variation of the modified functional at equilibrium is composed of the second variation of the standard reduced functional augmented with a new term, the second integral in Eq.~\eqref{eq:svar3}. It is particularly informative to compute this term before revealing the final result for the second variation. Using Eqs.~\eqref{eq:svar3}, \eqref{eq:pb_delD} and \eqref{eq:pb_dh-1}, this term is computed to be
\begin{eqnarray}\label{eq:pb_2term}
&&-2\int \delta_{h}h^{-1}(h(\phi^{*}))(\delta D[\phi])^{2} = 
4\,\frac{1}{4\pi}\int |\nabla\delta\phi|^{2} \\\nonumber
&&+ 2\beta C q^{2} \int e^{-\beta q \phi^{*}} (\delta\phi)^{2}
+ \frac{2}{\beta C q^{2}} \int e^{-\beta q \phi^{*}} 
\left(\frac{1}{4\pi}\nabla^{2}\delta\phi\right)^{2}.
\end{eqnarray}
We immediately see that, in the above expression, each of the three terms on the right-hand side of the equation are strictly non-negative. Moreover, a quick comparison with Eq.~\eqref{eq:pb_svarI} reveals that the first two terms on the right-hand side in the above equation not only cancel the negative-definite terms in Eq.~\eqref{eq:pb_svarI}, they lead to a net positive result.
Upon adding the right-hand sides of Eqs.~\eqref{eq:pb_svarI} and \eqref{eq:pb_2term}, we find that our modified functional has, at equilibrium, the following second variation:
\begin{eqnarray}
\delta^{2}F_{\phi}[\phi^{*}]&=&
\frac{3}{4\pi}\int |\nabla\delta\phi|^{2}
+ \beta C q^{2} \int e^{-\beta q \phi^{*}} (\delta\phi)^{2} \nonumber \\
&&+ \frac{2}{\beta C q^{2}} \int e^{-\beta q \phi^{*}} 
\left(\frac{1}{4\pi}\nabla^{2}\delta\phi\right)^{2},
\end{eqnarray}
which is positive-definite as desired.

\subsection{Magnetostatics}
The methods developed above for charged systems can be further extended to include magnetism and also suggest some avenues of investigation for electrodynamics. In this subsection, we consider the special case of magnetostatics. For the sake of simplicity, we focus on the case where the susceptibility $\mu$ is 1 everywhere, noting that the extension to the case of spatially-dependent $\mu$ can be carried out in similar fashion as the treatment of heterogeneous dielectric media in Sec.~\ref{sec:electrostatics_media}.

For the static magnetic field, we have the following starting functional:
\beq
I[\mathbf{B};\mathbf{A}]=\frac{1}{8\pi}\int\mathbf{B}\cdot\mathbf{B}
-\int\mathbf{A}\cdot\left(\frac{1}{4\pi}\mathbf{\nabla\times B} - \frac{1}{c}\mathbf{j}\right)
\eeq
where $c$ is the speed of light, $\mathbf{B}$ is the magnetic field, $\mathbf{j}$ is the current density, and $\mathbf{A}$ is the vector potential; the latter
obviously playing the role of a Lagrange multiplier. We note that in the static case we have $\nabla \cdot {\bf j}=0$. Variation of the functional with respect to the magnetic field leads to
\beq
\mathbf{B}=\mathbf{\nabla\times A}
\eeq
and the reduced functional of the potential is:
\begin{align}\label{eq:mag_IA}
I_{A}[\mathbf{A}]
&= \frac{1}{8\pi}\int\mathbf{\nabla\times A} \cdot \mathbf{\nabla\times A} \notag\\
&\quad\quad-\int \mathbf{A} \cdot \left( \frac{1}{4\pi} \mathbf{\nabla\times\nabla\times A} -\frac{1}{c} \mathbf{j}\right) \notag\\
&=-\frac{1}{8\pi}\int \mathbf{\nabla\times A}\cdot\mathbf{\nabla\times A}
+\frac{1}{c}\int\mathbf{A}\cdot\mathbf{j}
\end{align}
which is semi-negative-definite. It contains modes that can give a zero second functional derivative, but it is otherwise negative. In the electrostatic case, boundary conditions eliminated zero modes, but their elimination here is a bit more complicated. Any configuration of the form $\mathbf{A}=\nabla\eta$ does not contribute to the second functional derivative. If we impose the condition 
\beq\label{eq:gauge}
\nabla\cdot\mathbf{A}=0,
\eeq
the functional becomes negative-definite. We make this choice from now on. 

As before, the process of obtaining a positive-definite functional begins by recognizing the constraint 
\beq
D[\mathbf{A}] = \frac{1}{4\pi} \mathbf{\nabla\times\nabla\times A} -\frac{1}{c} \mathbf{j}
\eeq
from the first equality of Eq.~\eqref{eq:mag_IA}. Next, we employ the choice $h(\mathbf{A})=-(4\pi)^{-1}\mathbf{\nabla\times\nabla\times A}$, which 
following Eq.~\eqref{eq:gauge} becomes $h(\mathbf{A}) = (4\pi)^{-1}\nabla^{2}\mathbf{A}$.
We promptly compute the required inverse of the $h$ function:
$h^{-1}(\mathbf{v})= - \int G(\mathbf{x},\mathbf{y})\mathbf{v}(\mathbf{y})$.
The modified Lagrange multiplier for this problem follows from Eq.~\eqref{eq:modlm}:
\begin{align}
\mathbf{A}_{m} = - \int G 
\left(D[\mathbf{A}] -\frac{1}{4\pi}\mathbf{\nabla\times\nabla\times A}\right)
= \frac{1}{c}\int G \, \mathbf{j},
\end{align}
using which we obtain the modified functional
\begin{align}\label{eq:mag_FA}
F_{A}[\mathbf{A}]
&= \frac{1}{8\pi}\int\mathbf{\nabla\times A} \cdot \mathbf{\nabla\times A} \notag\\
&\quad\quad- \frac{1}{c}\iint G \, \mathbf{j} \cdot \left( \frac{1}{4\pi} \mathbf{\nabla\times\nabla\times A} -\frac{1}{c} \mathbf{j}\right).
\end{align}
A final reduction leads to the functional
\begin{align}
F_{A}[\mathbf{A}]
&=\frac{1}{8\pi}\int \mathbf{\nabla\times A}\cdot\mathbf{\nabla\times A}
+\frac{1}{4\pi c}\iint G \, \mathbf{j} \cdot \nabla^{2}\mathbf{A} \notag\\
&\quad\quad+ \frac{1}{c^{2}} \iint \mathbf{j}(\mathbf{x})\cdot G \, \mathbf{j}(\mathbf{y}).
\end{align}
Clearly, the above functional is a positive-definite functional of the vector potential when restricted to the subspace of divergence-less vector fields. 

\section{Possible extensions and applications}
After presenting these examples, we wish to make some general remarks on both the practical aspects of the use of the functionals we have developed, as well as about their application to other systems of interest.

First, we note that the modified functional has been obtained through a nonlinear transformation. As pointed out above, this reduces some of the range of applicability of the functional, at least as a single-valued real functional. Once we have achieved the goal of creating a positive-definite functional at the equilibrium point, one can hope that the exploration of non-equilibrium configurations never takes us to regions where the functional is not defined or it is negative-definite. There must always be a finite region, in some functional
topology or metric, where the positive-definiteness property remains valid. Thus, this becomes a practical implementation problem, but one that in principle has a solution.

Second, as the presented modified functionals retain a lot of the properties of the original reduced functionals, they clearly function as good alternatives to the latter. Their application to quantum mechanical problems may also be possible. In the path integral formulation of quantum mechanics, we explore non-stationary paths and weight them with the imaginary exponential of an action. In general, a modified action could provide the wrong weighting. However, if the constraint is a limit on the actually realizable paths; that is, if all the quantum fluctuations can only appear in the space that satisfies the constraint,
this approach would still in principle function and we look forward to develop such applications. Further, the method might also help in the elucidation of the properties of quantum systems where, for example, the ground state might be recovered as the solution of a variational problem.

Additionally, the rather ad-hoc nature of our modifications suggests that there might be other means of systematically modifying known functionals to improve their properties. Alternative approaches might include the use of different transformations of the multiplier, the iteration of the process of a single transformation, and the direct modification of the constraint.

As for other possible areas of application, we wish to note that since variational methods have long been applied to physical problems, there are likely plenty more systems in which the methodology might be of use. We note, for example, the extensive use of a constrained variational functional in investigations of shapes of membrane and elastic fibers \cite{guven1,guven2,fibers3}. We note below the relation with quantum field theory, but we also point out that other disciplines also work extensively with variational formulations, most notably applied control theory (see, for example, \cite{control}).

Finally, variational functionals have also been used in the discussion of field theories. Constrained functionals form an important and deeply investigated topic in the context of gauge field theories \cite{tyutin,henneaux}. We note however, that the motivation of our current approach, the finding of positive-definite functionals is not a relevant, nor likely possible, goal in relativistic electrodynamics and other field theories due to the signature of the Minkowski metric. Nevertheless, variants of this goal are likely of interest, for example, the construction of functionals with excitations that have non-negative mass. We will pursue such investigations in future publications.  

\begin{acknowledgments}
We thank G. Vernizzi for several valuable discussions. 
V.J. was funded by the Department of Defense Research and Engineering (DDR\&E) 
and the Air Force Office of Scientific Research (AFOSR) 
under Award No. FA9550-10-1-0167. 
F.J.S. was funded by the National Science Foundation (NSF) 
Grant No. DMR-1309027.
\end{acknowledgments}

\end{document}